\documentclass[letterpaper]{article}

\usepackage{aaai18}
\usepackage{booktabs}
\usepackage{multirow, bigstrut}
\usepackage{times}
\usepackage{hyperref}
\usepackage{algorithm}
\usepackage{algpseudocode}
\usepackage{tikz,pgf}
\usepackage{epsfig,amssymb,amsmath,ifthen,comment}
\usepackage{float}
\usepackage{subfigure}
\usepackage{color}
\usepackage{xr}
\newcommand{\E}{\mathbb{E}}
\newcommand{\I}{\mathbb{I}}

\newenvironment{thma}[1]{\par\noindent{\bf Theorem #1\ }\em}{\em}
\newenvironment{lema}[1]{\par\noindent{\bf Lemma #1\ }\em}{\em}

\title{Personalizing Path-Specific Effects}

\newenvironment{prf}{\noindent\textit{Proof:}\begin{mdseries}}{\end{mdseries}{\hfill\scriptsize$\Box$}} 

\DeclareMathOperator*{\argmax}{arg\,max}

\newtheorem{thm}{Theorem}
\newtheorem{lem}{Lemma}

\DeclareMathOperator{\dis}{dis}

\DeclareMathOperator{\pa}{pa}
\DeclareMathOperator{\de}{de}

\DeclareMathOperator{\ch}{ch}
\DeclareMathOperator{\an}{an}

\DeclareMathOperator{\nd}{nd}

\DeclareMathOperator{\sib}{sib}

\author{
}

\begin{document}

\maketitle

\begin{abstract}
Unlike classical causal inference, which often has an average
causal effect of a treatment within a population as a target, in settings
such as personalized medicine, the goal is to map a given unit's
characteristics to a treatment tailored to maximize the expected outcome
for that unit.  Obtaining high-quality mappings of this type is the goal
of the dynamic regime literature \cite{erica13statistical}, with connections
to reinforcement learning and experimental design.

Aside from the average treatment effects, mechanisms behind causal
relationships are also of interest.  A well-studied approach to mechanism
analysis is establishing average effects along with a particular set of causal
pathways, in the simplest case the direct and indirect effects.  Estimating
such effects is the subject of the mediation analysis literature \cite{robins92effects,pearl01direct}.

In this paper, we consider how unit characteristics may be used to
tailor a treatment assignment strategy that maximizes a particular
path-specific effect.  In healthcare applications, finding such a policy is
of interest if, for instance, we are interested in maximizing the chemical effect of a drug on an outcome (corresponding to the direct effect), while assuming drug adherence (corresponding to the indirect effect) is set to some reference level.

To solve our problem, we define counterfactuals associated with path-specific
effects of a policy, give a general identification algorithm for these counterfactuals, give a proof of completeness,
and show how classification algorithms in machine learning \cite{chen16owl}
may be used to find a high-quality policy.
We validate our approach via a simulation study.
\end{abstract}

\section{Introduction}

Establishing causal relationships between action and outcome is fundamental to
rational decision-making.  A gold standard for establishing causal relationships is
the randomized controlled trial (RCT), which may be used to establish average causal
effects within a population.  Causal inference is a branch of statistics that seeks
to predict effects of RCTs from observational data, where treatment assignment is
not randomized.  Such data is often gathered in observational studies, surveys given to
patients during follow up, and in hospital electronic medical records.

While RCTs and causal inference methods that predict results of hypothetical RCTs
establish whether a particular action is helpful \emph{on average},
optimal decision making must tailor decisions to specific situations.  In the context
of causal inference this involves finding a map between characteristics of an experimental
unit, such as baseline features, to an action that optimizes some outcome for that unit.
Methods for finding such maps are studied in the dynamic treatment regime literature,
and in off-policy reinforcement learning.

If an action is known to have a beneficial effect on some outcome, it is often desirable
to understand the causal mechanism behind this effect.  A popular type of mechanism
analysis is \emph{mediation analysis}, which seeks to decompose the average causal
effects into direct and indirect components, or more generally into components associated
with specific causal pathways.  These components of the average causal effect are known
as direct, indirect, and path-specific effects, and are also defined as a population average
\cite{robins92effects,pearl01direct,chen05ijcai}.
In this paper we introduce methods to personalize these types of effects, that is find mappings
from unit characteristics to actions that maximize some path-specific effect.


\subsection*{Why Personalize Path-Specific Effects?}

Just as it often makes sense to structure decision-making such that the
\emph{overall} effect of an action on the outcome is maximized for any specific unit,
in some cases it is appropriate to choose an action such that only a part of the effect
of an action on the outcome is maximized.  Consider management of care of HIV patients.
Since HIV is a chronic disease, care for HIV patients involves designing a long-term treatment
plan to minimize chances of viral failure (an undesirable outcome).  In designing such a plan, an important
choice is initiation of primary therapy, and a switch to a second line therapy.  Initiating or switching too early
risks unneeded side effects and "wasting" treatment efficacy, while initiating or switching too late risks viral
failure \cite{comparison06hernan}.

However, in the context of HIV, \emph{treatment adherence} is an important component of the overall
effect of the drug on the outcome.  Patients who do not take prescribed doses compromise the efficacy of
the drug, and different drugs may have different levels of adherence.  Thus, in HIV the overall effect of the
drug can be viewed as a combination of the chemical effect, and the adherence effect \cite{caleb17quantifying}.
Therefore, choosing an action that maximizes the overall effect of HIV treatment on viral failure entangles these two very different
mechanisms.  One approach to tailoring treatments to patients in a way that disentangles these mechanisms
is to find a policy that optimizes a part of the effect, say the chemical (direct) effect of the drug, while hypothetically
keeping the adherence levels to some reference level.  Finding such a policy yields information on how best to
assign drugs to maximize their chemical efficacy in settings where adherence levels can be controlled to a reference
level -- even if the only data available is one where patients have differential adherence.


\section*{Preliminaries}

We will consider causal models represented by acyclic directed graphs (DAGs), and acyclic directed mixed graphs
(ADMGs) representing classes of DAGs with hidden variables.  A DAG is a graph with directed ($\to$) edges with no
directed cycles, an ADMG is a graph with directed ($\to$) and bidirected ($\leftrightarrow$) edges with no directed cycles.

\subsection*{Graph Theory}

We will define statistical and causal models as sets of distributions defined by restrictions associated with graphs.  Thus we will
use vertices and variables interchangeably -- capital letters for a vertex or variable ($V$), bold capital letter for a set (${\bf V}$),
small letters for values ($v$), and bold small letters for sets of values (${\bf v})$.  For a set of values ${\bf a}$ of ${\bf A}$,
and a subset ${\bf A}^{\dag} \subseteq {\bf A}$, define ${\bf a}_{{\bf A}^{\dag}}$ to be a restriction of ${\bf a}$ to elements in
${\bf A}^{\dag}$.  
We will assume graphs with a vertex set ${\bf V}$.  The state space of $A$ will be
denoted by ${\mathfrak X}_A$, and the (Cartesian product) state space of ${\bf A}$ will be denoted by ${\mathfrak X}_{\bf A}$.

For a graph ${\cal G}$, and any $V \in {\bf V}$, we will define the following genealogic sets: parents,
children, ancestors, descendants, and siblings as:
 $\pa_{\cal G}(V) \equiv \{ W \in {\bf V} \mid W \to V \}$,
 $\ch_{\cal G}(V) \equiv \{ W \in {\bf V} \mid V \to W \}$,
 $\an_{\cal G}(V) \equiv \{ W \in {\bf V} \mid W \to \ldots \to V \}$,
 $\de_{\cal G}(V) \equiv \{ W \in {\bf V} \mid V \to \ldots \to W \}$.
 $\sib_{\cal G}(V) \equiv \{ W \in {\bf V} \mid V \leftrightarrow W \}$.
By convention, $\an_{\cal G}(V) \cap \de_{\cal G}(V) \cap \dis_{\cal G}(V) = \{ V \}$.
These sets generalize to ${\bf V}^{\dag} \subseteq {\bf V}$ disjunctively.  For example, $\pa_{{\cal G}}({\bf V}^{\dag}) \equiv
\bigcup_{V \in {\bf V}^{\dag}} \pa_{\cal G}(V)$.  For ${\bf A} \subseteq {\bf V}$, define $\pa^s_{\cal G}({\bf A}) \equiv
\pa_{\cal G}({\bf A}) \setminus {\bf A}$.

We define the set 
$\nd_{\cal G}(V) \equiv {\bf V} \setminus \de_{\cal G}(V)$.
The district of $V$ is defined as $\dis_{\cal G}(V) = \{ W \in {\bf V} \mid W \leftrightarrow \ldots \leftrightarrow V \}$.
The set of districts will be denoted by ${\cal D}({\cal G})$, and it always forms a partition of vertices in ${\cal G}$.
Given a graph ${\cal G}$ and ${\bf A} \subseteq {\bf V}$, denote by ${\cal G}_{\bf A}$ the subgraph of ${\cal G}$
containing only vertices in ${\bf A}$ and edges between these vertices.

\subsection*{Statistical And Causal Models Of A DAG}

A statistical model of a DAG or a Bayesian network, associated with a DAG ${\cal G}$,
is the set of distributions $p({\bf V})$ such that
$p({\bf V}) = \prod_{V \in {\bf V}} p(V \mid \pa_{\cal G}(V))$.
Such a $p({\bf V})$ is said to be Markov relative to ${\cal G}$.

Causal models of a DAG are also sets of distributions, but on counterfactual random variables.  Given
$Y \in {\bf V}$ and ${\bf A} \subseteq {\bf V} \setminus \{ Y \}$, a counterfactual variable, also known as
a potential outcome, and written as $Y({\bf a})$ represents variation in $Y$ in a hypothetical situation
where ${\bf A}$ were set to values ${\bf a}$ by an \emph{intervention operation} \cite{pearl09causality}.
Given a set ${\bf Y}$, define ${\bf Y}({\bf a}) \equiv \{ {\bf Y} \}({\bf a}) \equiv \{ Y({\bf a}) \mid Y \in {\bf Y} \}$.

Causal models of a DAG ${\cal G}$ can be viewed as modeling counterfactuals of the form $V({\bf a})$
where ${\bf a}$ are values of $\pa_{\cal G}(V)$.  These \emph{atomic counterfactuals} model the relationship
between $\pa_{\cal G}(V)$, representing direct causes of $V$, and $V$ itself.  From these, all other
counterfactuals may be defined using recursive substitution.  For any ${\bf A} \subseteq {\bf V} \setminus \{ V \}$,
{\small
\begin{align}
V({\bf a}) \equiv V({\bf a}_{\pa_{\cal G}(V) \cap {\bf A}},
\{ \pa_{\cal G}(V) \setminus {\bf A} \}({\bf a})).
\label{eqn:rec-sub}
\end{align}
}
A causal parameter is identified in a causal model if it is a function of the observed data distribution $p({\bf V})$.
In a causal model of a DAG ${\cal G}$, all interventional distributions\\
$p(\{ {\bf V} \setminus {\bf A} \}({\bf a}))$
are identified by the \emph{g-formula}:
{\small
\begin{align}
p(\{ {\bf V} \setminus {\bf A} \}({\bf a})) =
\!\!\!
\prod_{V \in {\bf V} \setminus {\bf A}}
\!\!\!
p(V | \pa_{{\cal G}}(V) \setminus {\bf A}, {\bf a}_{\pa_{\cal G}(V) \cap {\bf A}}).
\label{eqn:g}
\end{align}
}
Counterfactual responses to classical interventions are often compared on the mean difference scale for two values $a,a'$,
representing cases and controls: $\E[Y(a)] - \E[Y(a')]$.  This quantity is known as the average causal effect (ACE).

\subsection*{Edge Interventions}

A more general type of intervention is the \emph{edge intervention} \cite{shpitser15hierarchy}, which maps a set of directed edges in ${\cal G}$ to values of their source vertices.
We will write the mapping of a set of edges to values of their source vertices as the following shorthand: $(a_1W_1)_{\to}, (a_2W_2)_{\to}, \ldots, (a_kW_k)_{\to}$ to mean
that edge $(A_1W_1)_{\to}$ is assigned to value $a_1$, $(A_2W_2)_{\to}$ is assigned to value $a_2$, and so on until $(A_kW_k)_{\to}$ is assigned to value $a_k$.
Alternatively, we will write ${\mathfrak a}_{\alpha}$ to mean edges in $\alpha$ are mapped to values in the \emph{multiset} ${\mathfrak a}$ (since multiple edges
may share the same source vertex, and be assigned to different values).  For a subset $\beta \subseteq \alpha$, and an assignment ${\mathfrak a}_\alpha$
 denote ${\mathfrak a}_{\beta}$ to be a restriction of ${\mathfrak a}_{\alpha}$ to edges in $\beta$.

We will write counterfactual responses to edge interventions as $Y({\mathfrak a}_{\alpha})$ or, for a small set of edges, as: $Y((aY)_{\to},(a'M)_{\to})$ meaning the response to
$Y$ where $A$ is set to value $a$ for the purposes of the edge $(AY)_{\to}$ and to $a'$ for the purposes of the edge $(AM)_{\to}$.
An edge intervention that sets a set of edges $\alpha$ to values ${\mathfrak a}$ as $(a_1W_1)_{\to}, (a_2W_2)_{\to}, \ldots, (a_kW_k)_{\to}$ is defined via the following generalization of (\ref{eqn:rec-sub}):
{\small
\begin{align}
Y({\mathfrak a}_{\alpha}) \equiv Y({\mathfrak a}_{\{ (ZY)_{\to} \in \alpha \}},
\{  \pa^{\bar{\alpha}}_{\cal G}(Y) \}({\mathfrak a}_{\alpha})),
\label{eqn:rec-sub-e}
\end{align}
}
where $\pa^{\bar{\alpha}}_{\cal G}(Y) \equiv \{ W \in \pa_{\cal G}(Y) \mid (WY)_{\to} \not\in \alpha \}$.

Given ${\bf A}_{\alpha} \equiv \{ A \mid (AB)_{\to} \in \alpha \}$, and an edge intervention given by the mapping ${\mathfrak a}_{\alpha}$,
under the non-parametric structural equation model with independent errors (NPSEM-IE) of a DAG ${\cal G}$,
the joint distribution of the counterfactual responses $p(\{ V({\mathfrak a}_{\alpha}) \mid V \in {\bf V} \setminus {\bf A}_{\alpha} \})$ is identified,
via the \emph{edge g-formula}, which is the following generalization of (\ref{eqn:g}):
{\small
\begin{align}
\prod_{V \in {\bf V} \setminus {\bf A}_{\alpha}} p(V | {\mathfrak a}_{\{ (ZV)_{\to} \in \alpha \}}, \pa_{\cal G}^{\bar{\alpha}}(V)).
\label{eqn:rec-sub-e}
\end{align}
}
For example, in Fig \ref{fig:triangle} (a), $p(Y((aY)_{\to}, (a'M)_{\to})) = \sum_{W,M} p(Y \mid a, M, W) p(M \mid a', W) p(W)$.
This is sometimes known as the \emph{mediation formula} \cite{pearl11cmf}.

Counterfactual responses to edge interventions represent effects of treatments ${\bf A}$ along some but not all causal pathways.  In simplest cases,
these responses can be used, often on the mean difference scale, to define direct and indirect effects \cite{robins92effects},\cite{pearl01direct}.
For example, in the model given by the DAG in Fig \ref{fig:triangle} (a), the direct effect of $A$ on $Y$ is defined as
$\E[Y((aY)_{\to},(aM)_{\to})] - \E[Y((a'Y)_{\to},(aM)_{\to})]$ which is equal to
$\E[Y(a)] - \E[Y(a',M(a))]$.  The indirect effect may be defined similarly as
$\E[Y((a'Y)_{\to},(aM)_{\to})] - \E[Y((a'Y)_{\to},(a'M)_{\to})]$, which is equal to
$\E[Y(a',M(a))] - \E[Y(a')]$.  The direct and indirect effects defined in this way
add up to the ACE.

Edge interventions represent a special case of a more general notion of a \emph{path-specific effect} \cite{pearl01direct} which, unlike path-specific
effects, happens to always be identified under an NPSEM-IE of a DAG, via (\ref{eqn:rec-sub-e}).  Path-specific effects may not be identified
even in a DAG model, due to the presence of a \emph{recanting witness} \cite{chen05ijcai}.

\subsection*{Responses To Treatment Policies}

In settings such as personalized medicine, counterfactual responses to conditional interventions that set treatment values in response to other variables via
a known function are 
often of interest.  Given a DAG ${\cal G}$, a topological ordering $\prec$, and a set ${\bf A} \subseteq {\bf V}$, for each
$A \in {\bf A}$, define ${\bf W}_A$ to be some subset of predecessors of $A$ according to $\prec$.  Then, given a set of functions ${\bf f}_{\bf A}$
of the form $f_A : {\mathfrak X}_{{\bf W}_A} \mapsto {\mathfrak X}_A$, we define $Y({\bf f}_{\bf A})$, the counterfactual response $Y \in {\bf V}$ to ${\bf A}$
being intervened on according to ${\bf f}_{\bf A} \equiv \{ f_A \mid A \in {\bf A} \}$, as
{\small
\begin{align}
Y
(
\{ f_A({\bf W}_A({\bf f}_{\bf A}) | A \in \pa_{\cal G}(Y) \cap {\bf A} \}), \{ \pa_{\cal G}(Y) \setminus {\bf A} \}({\bf f}_{\bf A})
).
\label{eqn:rec-sub-f}
\end{align}
}

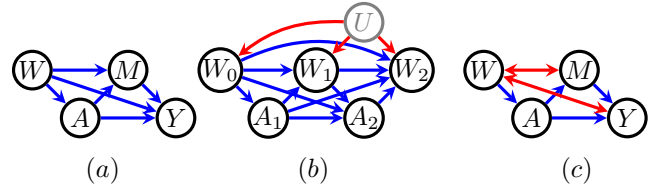
\begin{figure}
\begin{center}
\begin{tikzpicture}[>=stealth, node distance=0.9cm]
    \tikzstyle{format} = [draw, very thick, circle, minimum size=5.0mm,
	inner sep=0pt]

	\begin{scope}
		\path[->, very thick]
			node[format] (w) {$W$}
			node[format, below right of=w] (a) {$A$}
			node[format, above right of=a] (m) {$M$}
			node[format, below right of=m] (y) {$Y$}

			(w) edge[blue] (a)
			(a) edge[blue] (y)
			(a) edge[blue] (m)
			(m) edge[blue] (y)
			(w) edge[blue] (m)
			(w) edge[blue] (y)

			node[below of=a, yshift=0.2cm, xshift=0.3cm] (l) {$(a)$}
		;
	\end{scope}

%
%
%
	\begin{scope}[xshift=2.5cm]
		\path[->, very thick]
			node[format] (w) {$W_0$}
			node[format, below right of=w] (a) {$A_1$}
			node[format, above right of=a] (m) {$W_1$}
			node[format, gray, above right of=m] (u) {$U$}
			node[format, below right of=m] (z) {$A_2$}
			node[format, above right of=z] (y) {$W_2$}

			(u) edge[red, bend right=20] (w)
			(w) edge[blue] (a)
			(a) edge[blue] (z)
			(a) edge[blue] (m)
			(m) edge[blue] (z)
			(w) edge[blue] (m)
			(w) edge[blue] (z)
			(z) edge[blue] (y)
			(m) edge[blue] (y)
			(a) edge[blue] (y)
			(w) edge[blue, bend left=25] (y)
			(u) edge[red] (m)
			(u) edge[red] (y)

			node[below of=a, yshift=0.2cm, xshift=0.6cm] (l) {$(b)$}
		;
	\end{scope}

	\begin{scope}[xshift=6.0cm]
		\path[->, very thick]
			node[format] (w) {$W$}
			node[format, below right of=w] (a) {$A$}
			node[format, above right of=a] (m) {$M$}
			node[format, below right of=m] (y) {$Y$}

			(w) edge[blue] (a)
			(a) edge[blue] (y)
			(a) edge[blue] (m)
			(m) edge[blue] (y)
			(w) edge[<->,red] (m)
			(w) edge[<->,red] (y)

			node[below of=a, yshift=0.2cm, xshift=0.6cm] (l) {$(c)$}
		;
	\end{scope}

%
%
\end{tikzpicture}
\end{center}
\caption{(a) A simple causal DAG, with a single treatment $A$, a single outcome $Y$, a vector $W$ of baseline variables, and a single mediator $M$.
(b) A more complex causal DAG with two treatments $A_1,A_2$, an intermediate outcome $W_1$, and the final outcome $W_2$.
(c) A graph where $p(Y(a,M(a')))$ is identified, but $p(Y(f_A(W),M(a)))$ is not.
}
\label{fig:triangle}
\end{figure}

As an example, in the graph in Fig. \ref{fig:triangle} (b), if we are interested in evaluating the efficacy of a policy set $\{ f_{A_1} : \mathfrak{X}_{W_0} \mapsto \mathfrak{X}_{A_1},
f_{A_2} : \mathfrak{X}_{\{W_0,W_1\}} \mapsto \mathfrak{X}_{A_2} \}$ as far as their effect on the outcome $W_2$, we could evaluate it via the random variable
$Y(f_{A_1}, f_{A_2})$ defined as
{\scriptsize
$W_2
(
f_{A_2}(W_1(f_{A_1}(W_0),W_0), W_0),
W_1(f_{A_1}(W_0),W_0), f_{A_1}(W_0), W_0
)$.
}
The efficacy of a particular set of policies may be evaluated on the mean scale as
$\E[Y(f_{A_1}, f_{A_2})]$.
In a causal model of a DAG, given any policy set, the effect of ${\bf f}_{\bf A}$ on ${\bf V} \setminus {\bf A}$,
represented by the distribution $p(\{ V({\bf f}_{\bf A}) \mid V \in {\bf V} \setminus {\bf A}) \})$,
is identified by the following modification of (\ref{eqn:g}):
{\small
\begin{align}
\prod_{V \in {\bf V} \setminus {\bf A}}
\!\!\!
\!\!\!
p
(
V |
\{ f_A({\bf W}_A({\bf f}_{\bf A})) | A \!\in\! {\bf A} \!\cap\! \pa_{\cal G}(V) \}, \{ \pa_{\cal G}(V) \!\setminus\! {\bf A} \}({\bf f}_{\bf A})
).
\end{align}
}
For example, $p(Y(f_A,f_Z))$ is identified as
{\small
\begin{align}
\notag
\sum_{W_0,W_1}
p(W_2 | W_0, f_{A_1}(W_0), W_1, f_{A_2}(W_0,W_1))\times \\
p(W_1 | W_0, f_{A_1}(W_0)) p(W_0).
\label{eqn:id-f-dag}
\end{align}
}

\section*{Identification In Hidden Variable DAGs}

In a causal model of a DAG where some variables are unobserved, not every causal parameter is identifiable, that is
not every parameter is a function of the observed data distribution.
Given a DAG ${\cal G}$ with a vertex set ${\bf V} \cup {\bf H}$, where ${\bf V}$ are observed, and ${\bf H}$ are hidden,
define a \emph{latent projection} ${\cal G}({\bf V})$ to be an ADMG with observed variables ${\bf V}$ with an edge
$(AB)_{\to}$ if there exists a directed path from $A$ to $B$ in ${\cal G}$ with all intermediate vertices in ${\bf H}$, and
an edge $(AB)_{\leftrightarrow}$ if there exists a path without consecutive edges $\to \circ \gets$ from $A$ to $B$ with
the first edge on the path of the form $A \gets$ and the last edge on the path of the form $\to B$.  A variable pair in
${\cal G}({\bf V})$ may be connected by both a directed and a bidirected edge.
It is known that all hidden variable DAGs which share latent projections share identification theory.  Thus, we will describe
identification results on latent projection ADMGs directly.  General algorithms for identification of interventional distributions
were given in \cite{tian02on},\cite{shpitser06id}, for responses to edge interventions in \cite{shpitser13cogsci}, and for policies in
\cite{tian08dynamic}.  Here we reformulate these results via simple one line formulas using conditional ADMGs and
a fixing operator.

\subsection*{Conditional ADMGs, Kernels, And Fixing}

A conditional ADMG (CADMG) ${\cal G}({\bf V},{\bf W})$ is an ADMG where ${\bf W}$ are \emph{fixed vertices} with the additional property that
for all $W \in {\bf W}$, $\sib_{\cal G}(W) \cap \pa_{\cal G}(W) = \emptyset$.  A kernel $q_{\bf V}({\bf V} \mid {\bf W})$ is a mapping
from ${\mathfrak X}_{\bf W}$ to normalized densities over ${\bf V}$.  A conditional distribution is one type of kernel, but others are possible.
Conditioning and marginalization are defined in kernels in the usual way.  For ${\bf A} \subseteq {\bf V}$,
{\small
\begin{align*}
q_{\bf V}({\bf A} | {\bf W}) &\equiv \sum_{{\bf V} \setminus {\bf A}} q_{\bf V}({\bf V} | {\bf W}); \hspace{1pt}
q_{\bf V}({\bf V} \setminus {\bf A} | {\bf A} \cup {\bf W}) \equiv \frac{q_{\bf V}({\bf A} | {\bf W})}{q_{\bf V}({\bf A} | {\bf W})}.
\end{align*}
}
For $A \in {\bf V}$, a fixing operator $\phi_A({\cal G}({\bf V},{\bf W}))$ produces a new CADMG ${\cal G}({\bf V} \setminus \{A\},{\bf W} \cup\{A\})$,
where all edges into $A$ are removed.
For a CADMG ${\cal G}({\bf V},{\bf W})$ and kernel $q_{\bf V}({\bf V}\mid{\bf W})$, and $A \in {\bf V}$,
a fixing operator $\phi_A(q_{\bf V};{\cal G})$ produces a new kernel
$\tilde{q}_{{\bf V} \setminus\{A\}}({\bf V}\setminus\{A\}\mid{\bf W}\cup\{A\}) \equiv
q_{\bf V}({\bf V} \mid {\bf W})
/q_{\bf V}(A \mid {\bf W} \cup \nd_{\cal G}(A))
$.

A sequence $\langle A_1, A_2, \ldots, A_k \rangle$ of vertices in ${\bf V}$ is said to be fixable if $A_1$ is fixable in ${\cal G}$,
$A_2$ is fixable in $\phi_{A_1}({\cal G})$, and so on, with $A_k$ being fixable in $\phi_{A_{k-1}}(\ldots \phi_{A_2}(\phi_{A_1}({\cal G})) \ldots )$.
A consequence of a theorem in \cite{richardson17nested} states that if $p({\bf H} \cup {\bf V})$ is Markov relative to ${\cal G}({\bf H} \cup {\bf V})$,
for any two sequences $\langle A_{i_1}, \ldots, A_{i_k} \rangle$, $\langle A_{j_1}, \ldots, A_{j_k} \rangle$ fixable in ${\cal G}({\bf V})$,
graphs and kernels obtained from applying these sequences to ${\cal G}({\bf V})$ and $p({\bf V})$ are the same.
For this reason, we will consider fixable sets.  A set is fixable in ${\cal G}$ if it is possible to arrange its elements into a fixable sequence.
All sequences are fixable in a DAG.  For ${\bf A}$ fixable in ${\cal G}$, we will define $\phi_{\bf A}(.)$, applied to either graphs or kernels,
to be a composition of $\phi$ applied in order to some fixable sequence of elements in ${\bf A}$.
If ${\bf A} \subseteq {\bf V}$ is fixable in ${\cal G}$, then the set ${\bf V} \setminus {\bf A}$ is called a reachable set.

Given a kernel $q_{\bf V}({\bf V} \mid {\bf W}) \equiv \phi_{{\bf W}}(p({\bf V} \cup {\bf W}); {\cal G}({\bf V} \cup {\bf W}))$, and given ${\bf a} \in {\mathfrak X}_{\bf A}$,
for ${\bf A} \subseteq {\bf W}$, define $ \phi^{\bf a}_{{\bf W}}(p({\bf V} \cup {\bf W}); {\cal G}({\bf V} \cup {\bf W}))$ to be a kernel
$\tilde{q}_{\bf V}({\bf V} \mid {\bf W} \setminus {\bf A})$ such that for any ${\bf w} \in {\mathfrak X}_{{\bf W} \setminus {\bf A}}$,
$\tilde{q}_{\bf V}({\bf V} \mid {\bf w}) = q_{\bf V}({\bf V} \mid {\bf w},{\bf a})$.

\subsection*{Identification Algorithms Via The Fixing Operator}

A complete algorithm for identifying interventional distributions of the form
$p({\bf Y}({\bf a}))$ for ${\bf Y} \subseteq {\bf V} \setminus {\bf A}$ was given in
\cite{tian02on}.  This algorithm can be rephrased using the fixing operator as follows.
Let ${\bf Y}^* \equiv \an_{{\cal G}_{{\bf V} \setminus {\bf A}}}({\bf Y})$.
Then if for every ${\bf D} \in {\cal D}({\cal G}_{{\bf Y}^*})$, ${\bf D}$ is reachable in ${\cal G}$,
then for ${\bf Y} \subseteq {\bf V} \setminus {\bf A}$,
{\small
\begin{align}
p({\bf Y}({\bf a})) =
\sum_{{\bf Y}^* \setminus {\bf Y}} \prod_{{\bf D} \in {\cal D}({\cal G}_{{\bf Y}^*})}
\phi^{{\bf a}_{\pa^s_{\cal G}({\bf D})}}_{{\bf V} \setminus {\bf D}}(p({\bf V}); {\cal G}).
\label{eqn:id}
\end{align}
}
If some ${\bf D} \in {\cal D}({\cal G}_{{\bf Y}^*})$ is not reachable in ${\cal G}$,
$p({\bf Y}({\bf a}))$ is not identifiable.
See theorem 60 in \cite{richardson17nested}. 

Identification of path-specific effects where each path is associated with one of two
possible value sets ${\bf a},{\bf a}'$ was given a general characterization in
\cite{shpitser13cogsci} via the \emph{recanting district criterion}.
Here, we reformulate this result in terms of the fixing operator in a way that generalizes (\ref{eqn:id}), and
applies to the response of any edge intervention, including those that set edges to multiple values rather than
two.  This result can also be viewed as a generalization of \emph{node consistency} of edge interventions in DAG models,
found in \cite{shpitser15hierarchy}.

Given ${\bf A}_{\alpha} \equiv \{ A \mid (AB)_{\to} \in \alpha \}$, and an edge intervention given by
the mapping ${\mathfrak a}_{\alpha}$, define ${\bf Y}^* \equiv \an_{{\cal G}_{{\bf V} \setminus {\bf A}_{\alpha}}}({\bf Y})$.
The joint distribution of the counterfactual response
$p(\{ {\bf V} \setminus {\bf A}_{\alpha} \}({\mathfrak a}_{\alpha}))$
is identified, under the NPSEM-IE, if and only if 
$p(\{ {\bf V} \setminus {\bf A}_{\alpha} \}({\bf a}))$
is identified via
(\ref{eqn:id}), and for every ${\bf D} \in {\cal D}({\cal G}_{{\bf Y}^*})$, for every $A \in {\bf A}_{\alpha}$,
either every directed edge out of $A$ into ${\bf D}$ is in $\alpha$ and ${\mathfrak a}_{\alpha}$ agrees on value assignments
to those edges, or every directed edge out of $A$ into ${\bf D}$ is not in $\alpha$.
\begin{thm}
Under above assumptions,
$p(\{ {\bf V} \setminus {\bf A}_{\alpha} \}({\mathfrak a}_{\alpha}))$ is 
{\small
\begin{align}
\sum_{{\bf Y}^* \setminus {\bf Y}} \prod_{{\bf D} \in {\cal D}({\cal G}_{{\bf Y}^*})}
\phi
^{{\mathfrak a}_{ \{ (AD)_{\to} \in \alpha \mid D \in {\bf D}, A \not\in {\bf D} \} }}
_{{\bf V} \setminus {\bf D}}(p({\bf V}); {\cal G}).
\label{eqn:id-e}
\end{align}
}
\label{thm:id-e}
\end{thm}

A general algorithm for identification of responses to a set of policies ${\bf f}_{\bf A}$ is given
in \cite{tian08dynamic}.  We again reformulate this algorithm in terms of the fixing operator.
Define a graph ${\cal G}_{{\bf f}_{\bf A}}$ to be a graph obtained from ${\cal G}$ by removing all
edges into ${\bf A}$, and adding for any $A \in {\bf A}$, directed edges from ${\bf W}_A$ to $A$.
Define ${\bf Y}^* \equiv \an_{{\cal G}_{{\bf f}_{\bf A}}}({\bf Y}) \setminus {\bf A}$.  Then
$p({\bf Y}({\bf f}_{\bf A}))$
is identified in ${\cal G}$ if
$p({\bf Y}^*({\bf a}))$
is identified.
Moreover, the identification formula is
{\small
\begin{align}
\sum_{({\bf Y}^* \cup {\bf A}) \setminus {\bf Y}}
\prod_{{\bf D} \in {\cal D}({\cal G}_{{\bf Y}^*})}
\phi^{\tilde{\bf a}_{\pa^s_{\cal G}({\bf D}) \cap {\bf A}}}_{{\bf V} \setminus {\bf D}}(p({\bf V}); {\cal G}),
\label{eqn:id-f}
\end{align}
}
where $\tilde{\bf a}_{\pa^s_{\cal G}({\bf D}) \cap {\bf A}}$ is defined to be
$\{ A = f_A({\bf W}_A) \mid A \in \pa_{\cal G}({\bf D}) \cap {\bf A} \}$
if $\pa_{\cal G}({\bf D}) \cap {\bf A}$ is not empty, and is defined to be the
empty set otherwise.  The sum over ${\bf A}$ is vacuous if ${\bf f}_{\bf A}$
is a set of deterministic policies.

\section*{Path-Specific Policies}

Fix a set of directed edges $\alpha$, and define ${\bf A}_{\alpha} \equiv \{ A \mid (AB)_{\to} \in \alpha \}$ as before.
Denote $\bar{\alpha}$ to be the set of outgoing edges from elements in ${\bf A}_{\alpha}$ not in $\alpha$.
Consider a set $\{ {\bf W}_A \mid A \in {\bf A}_{\alpha} \}$ defined as before with respect to a topological ordering $\prec$.
We are going to consider a simple version of path-specific policies where for variables in ${\bf A}$ we wish to
intervene on, all outgoing edges for every $A \in {\bf A}$ are either associated with a reference
policy $f'_{A} : {\mathfrak X}_{{\bf W}_A} \mapsto {\mathfrak X}_A$
(for edges not in $\alpha$), or a policy of interest $f_{A} : {\mathfrak X}_{{\bf W}_A} \mapsto {\mathfrak X}_A$
(for edges in $\alpha$). Our results generalize to more complex types of path-specific policies, but we do not pursue this
here in the interests of space.  Generally, we will let $f'_{A}$ be a simple policy that sets $A$ to a reference value $a$,
ignoring ${\bf W}_A$.  Such reference policies are the most relevant in practice. 

We now define counterfactual responses to these types of policies, which we denote by
$(\tilde{\bf f}_{\bf A})_{\alpha,\overline{\alpha}} =
({\bf f}_{\bf A})_{\alpha},({\bf f}'_{\bf A})_{\overline{\alpha}}$,
where ${\bf f}_{\bf A} \equiv \{ f_A \mid A \in {\bf A} \}$, ${\bf f}'_{\bf A} \equiv \{ f'_A \mid A \in {\bf A} \}$, $\tilde{\bf f}_{\bf A} = {\bf f}_{\bf A} \cup {\bf f}'_{\bf A}$,
and the subscripts $\alpha,\overline{\alpha}$
are meant to denote that these policies only apply for the purposes of those respective edge sets.
Define
$Y((\tilde{\bf f}_{\bf A})_{\alpha,\overline{\alpha}})$
as
{\small
\begin{align*}
Y(
\{ f'_A({\bf W}_A((\tilde{\bf f}_{\bf A})_{\alpha,\overline{\alpha}}) | (AY)_{} \in \overline{\alpha} \},\\
\{ f_A({\bf W}_A((\tilde{\bf f}_{\bf A})_{\alpha,\overline{\alpha}}) | (AY)_{} \in {\alpha} \},
\{ W((\tilde{\bf f}_{\bf A})_{\alpha,\overline{\alpha}}) | (WY)_{} \not\in \alpha,\overline{\alpha} \}
)
\end{align*}
}
This definition generalizes both (\ref{eqn:rec-sub-e}) and (\ref{eqn:rec-sub-f}) in an appropriate way.
As an example, in Fig. \ref{fig:triangle} (a), a policy $f_A(W)$ that sets $A$ to a value only with respect to the edge
$(AY)_{\to}$, and a reference value $a$ that $A$ assumes with respect to the edge $(AM)_{\to}$ results in the
counterfactual $Y(f_A(W), M(a,W), W)$.
In the graph in Fig. \ref{fig:triangle} (b), the response of $W_2$ to $A_1,A_2$ being set according to $f_{A_1}(W_0), f_{A_2}(W_1,A_1,W_0)$
with respect to $(A_1W_2)_{\to}$, $(A_2W_2)_{\to}$, and set to $a_1,a_2$ for all other edges, is
{\small
\begin{align}
W_2(
	f_{A_2}(W_1(a_1,W_0), a_1, W_0),
	W_1(a_1,W_0),
	f_{A_1}(W_0),
	W_0
	).
\label{eqn:cntfl}
\end{align}
}

\subsection*{Identification Of 
Path-Specific Policies}

Having condensed existing identification results on responses to policies (\ref{eqn:id-f}) and responses to edge interventions
arising in mediation analysis (\ref{eqn:id-e}), we generalize these results to give an identification result for responses to
path-specific policies, via the following theorem


\begin{thm}
Define ${\cal G}_{\tilde{\bf f}_{{\bf A}_{\alpha}}}$ as ${\cal G}_{{\bf f}_{\bf A}}$ before, and let
${\bf Y}^* \equiv \an_{{\cal G}_{\tilde{\bf f}_{{\bf A}_{\alpha}}}}({\bf Y}) \setminus {\bf A}_{\alpha}$.
Then
$p({\bf Y}((\tilde{\bf f}_{\bf A})_{\alpha,\overline{\alpha}}))$
is identified if
$p({\bf Y}^*({\bf a}))$
is identified, and for every ${\bf D} \in {\cal D}({\cal G}_{{\bf Y}^*})$,
and every $A \in {\bf A}_{\alpha}$, either every directed edge out of $A$ into ${\bf D}$ is in $\alpha$,
or every directed edge out of $A$ into ${\bf D}$ is not in $\alpha$.  Moreover, the identifying formula is
{\small
\begin{align}
\sum_{({\bf Y}^* \cup {\bf A}_{\alpha}) \setminus {\bf Y}} \prod_{{\bf D} \in {\cal D}({\cal G}_{{\bf Y}^*})}
	\phi^{\tilde{\bf a}_{\pa^s_{\cal G}({\bf D}) \cap {\bf A}_{\alpha}}}_{{\bf V} \setminus {\bf D}}(p({\bf V});{\cal G}),
\label{eqn:id-ef}
\end{align}
}
where ${\tilde{\bf a}_{\pa^s_{\cal G}({\bf D}) \cap {\bf A}_{\alpha}}}$ is defined to be
$\{ A = f_A({\bf W}_A) \mid A \in \pa^{\alpha}_{\cal G}({\bf D}) \cap {\bf A}_{\alpha} \} \cup
\{ A = f'_A({\bf W}_A) \mid A \in \pa^{\overline{\alpha}}_{\cal G}({\bf D}) \cap {\bf A}_{\alpha} \} $,
if $\pa_{\cal G}({\bf D}) \cap {\bf A}_{\alpha}$
is not empty, and is defined to be the empty set otherwise.
\label{thm:id-ef}
\end{thm}
Responses to path-specific policies are identified in strictly fewer cases compared to responses to edge interventions.
This is because ${\bf Y}^*$ is a larger set in the former case.  As an example, consider the graph in Fig. \ref{fig:triangle} (c),
where we are interested either in the counterfactual $p(Y(a,M(a')))$, used to define pure direct effects, and the counterfactual
$p(Y(f_A(W),M(a')))$.

For the former, we have ${\bf Y}^* = \{ Y,M\}$, and $p(Y(a,M(a')))$ equal to
{\small
\begin{align*}
\sum_m \left( \frac{\sum_w p(Y,m | a,w) p(w)}{ \sum_w p(m \mid a,w)p(w)} \right)
\sum_{w} p(m \mid a',w) p(w)
\end{align*}
}
We omit the detailed derivation 
in the interests of space.

For the latter, however, ${\bf Y}^* = \{ Y,M,W\}$, and since $\alpha = \{ (AY)_{\to} \}$ and $\bar{\alpha} = \{ (AM)_{\to} \}$,
Theorem \ref{thm:id-ef} is insufficient to conclude identification.
An example where identification is possible is shown in Fig. \ref{fig:triangle} (b).  Here, we are interested in optimizing the direct
effect of $A_1$ and $A_2$ on $W_2$ via policies $f_{A_1}(W_0)$ and $f_{A_2}(W_1,W_0)$, while keeping the indirect effect of
$A_1$ on $W_2$ through $W_1$ at a reference level $W_1(a_1)$.  This yields the counterfactual (\ref{eqn:cntfl}), which is identified as
{\small
\[
\sum_{W_0,W_1}
\!\!\!
p(W_2 | f_{A_2}(W_1,W_0), W_1, f_{A_1}(W_0), W_0) 
p(W_1 | a_1, W_0) p(W_0).
\]
}

\begin{figure}
\begin{center}
\begin{tikzpicture}[>=stealth, node distance=0.9cm]
    \tikzstyle{format} = [draw, very thick, circle, minimum size=5.0mm,
	inner sep=0pt]

	\begin{scope}[xshift=0.0cm]
		\path[->, very thick]
			node[format] (a1) {$A_1$}
			node[format, above of=a1] (w0) {$W_0$}
			node[format, right of=w0] (m1) {$M_1$}
			node[format, right of=a1] (w1) {$W_1$}
			node[format, above right of=w1, yshift=-0.2cm, xshift=0.2] (w2) {$W_2$}
			
			(w0) edge[blue] (a1)
			(a1) edge[blue] (m1)
			(m1) edge[blue] (w1)
			(w1) edge[blue] (w2)
			
			(w0) edge[blue] (m1)
			(w0) edge[blue] (w1)
			(w0) edge[blue, bend left=60] (w2)
			
			(m1) edge[blue] (w2)
			(a1) edge[blue] (w1)
			(a1) edge[<->, red, bend right=60] (w2)
			
			(w0) edge[<->, red, bend left=30] (m1)
			
			node[below of=m1, yshift=-0.7cm, xshift=0.0cm] (l) {$(a)$}
			;
	\end{scope}

	\begin{scope}[xshift=2.5cm]
		\path[->, very thick]
			node[] (a1) {}
			node[format, above of=a1] (w0) {$W_0$}
			node[format, right of=w0] (m1) {$M_1$}
			node[format, right of=a1] (w1) {$W_1$}
			node[format, above right of=w1, yshift=-0.2cm, xshift=0.2] (w2) {$W_2$}
			
			(m1) edge[blue] (w1)
			(w1) edge[blue] (w2)
			
			(w0) edge[blue] (m1)
			(w0) edge[blue] (w1)
			(w0) edge[blue, bend left=60] (w2)
			
			(m1) edge[blue] (w2)
			
			(w0) edge[<->, red, bend left=30] (m1)
			
			node[below of=m1, yshift=-0.7cm, xshift=0.0cm] (l) {$(b)$}
			;
	\end{scope}
	
	\begin{scope}[xshift=5.0cm]
		\path[->, very thick]
			node[format] (w0) {$W_0$}
			node[format, below right of=w0] (a1) {$A_1$}
			node[format, above right of=w0] (m1) {$M_1$}
			node[format, below right of=m1] (w1) {$W_1$}

			node[format, gray, above right of=m1] (u) {$U$}
			
			node[format, below right of=w1] (a2) {$A_2$}
			node[format, above right of=w1] (m2) {$M_2$}
			node[format, below right of=m2] (w2) {$W_2$}

			(w0) edge[blue] (a1)
			(w0) edge[blue] (m1)
			(w0) edge[blue] (w1)

			(a1) edge[blue] (m1)
			(a1) edge[blue] (w1)
			
			(m1) edge[blue] (w1)

			(w1) edge[blue] (a2)
			(w1) edge[blue] (m2)
			(w1) edge[blue] (w2)

			(a2) edge[blue] (m2)
			(a2) edge[blue] (w2)
			
			(m2) edge[blue] (w2)

			(u) edge[red, bend right=35] (w0)
			(u) edge[red, bend right=0] (w1)
			(u) edge[red, bend left=35] (w2)
			
			(m1) edge[blue] (m2)
			(a1) edge[blue] (a2)
			
			(a1) edge[blue, bend right=0] (w2)

			node[below of=a1, yshift=0.2cm, xshift=0.6cm] (l) {$(c)$}
		;
	\end{scope}

\end{tikzpicture}
\end{center}
\caption{
(a) A causal model where $p(W_2(\{ f_{A_1}(W_0) \}_{(A_1W_1)_{\to}}, a_{(A_1M_1)_{\to}}))$ is identified.
(b) The graph ${\cal G}_{{\bf Y}^*}$, where ${\bf Y}^* = \{ W_0, M_1, W_1, W_2 \}$ obtained from (a).
(c) A causal model representing the chemical effect of HIV mediation, and adherence on viral failure.
}
\label{fig:med_long}
\end{figure}
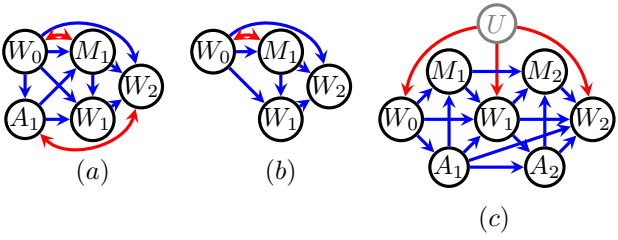

Generalizations of the example in Fig. \ref{fig:triangle} (b) are the most relevant in practice, as their causal structure corresponds to longitudinal
observational studies, of the kind considered in \cite{robins86new}, and many other papers.  However, we illustrate complications that may arise
in identifiability of responses to path-specific policies with the following, more complex, example in Fig. \ref{fig:med_long} (a).
Here, the distribution $p(W_2(\{ f_{A_1}(W_0) \}_{(A_1W_1)_{\to}}, a_{(A_1M_1)_{\to}}))$
is identified via: 
{\small
\begin{align*}
\sum_{W_0,A_1,M_1,W_1}
\!\!\!\!\!
\!\!\!\!\!
\begin{array}{c}
\left[
p(W_1 \mid M_1, f_{A_1}(W_0), W_0)
\right]
\left[
p(M_1 \mid a_1, W_0) p(W_0)
\right]\times\\
\left[
\sum_{W_0,A_1} 
p(W_2 \mid W_1,M_1,A_1,W_0) p(A_1,W_0)
\right],
\end{array}
\end{align*}
}
with the detailed derivation given in the Appendix.
In this example, the graph entails an identifying formula that does \emph{not} resemble a factorization where a conditional distribution
of the treatment $A_1$ is replaced by a policy $f_{A_1}$, such as (\ref{eqn:id-f-dag}).
Instead, due to the presence of a bidirected arrow connecting $A_1$ and $W_2$,
the identifying functional resembles the functional arising from the front-door criterion \cite{pearl09causality}.

\section*{On Completeness 
}

The ID algorithm phrased in terms of (\ref{eqn:id}) via the fixing operator is known to be \emph{complete} for non-parametric identification
\cite{huang06do,shpitser06id}.  Completeness here means that failure of identification means no other method is able to yield
identification under the same model.  A similar result does not, to the authors' knowledge, exist for the identification algorithm for
responses to policies in \cite{tian08dynamic}, and rephrased as (\ref{eqn:id-f}).

Here we give an argument for completeness of algorithms corresponding to (\ref{eqn:id-f}), and (\ref{eqn:id-ef}) under the
assumption that not only is the causal model non-parametric, but the set of policies ${\bf f}_{\bf A}$ consists of arbitrary
functions.  In other words, to show non-identifiability it will suffice to exhibit two \emph{unrestricted} elements in the causal model,
and \emph{any set} of functions ${\bf f}_{\bf A}$ such that the two elements agree on $p({\bf V})$ but disagree on
$p({\bf Y}({\bf f}_{\bf A}))$.
If the set of policies of interest ${\bf f}_{\bf A}$ is restricted, or alternatively if the causal model has parametric restrictions,
completeness may no longer hold.  To see this, consider Fig. \ref{fig:triangle} (c) where we pick functions $f_A(W)$ that sets $A$ to $a$ for the purposes of
$(AY)_{\to}$, and $f'_A(W)$ that sets $A$ to $a'$ for the purposes of $(AM)_{\to}$.  In other words, both functions ignore $W$.
In this restricted class, $p(Y((f_A)_{(AY)_{\to}}, (f'_A)_{(AM)_{\to}}))$ is in fact identifiable, since this distribution is equal to
$p(Y(a,M(a')))$, which was shown to be identifiable in the previous section.
We defer all proofs to the Appendix.


\begin{lem}
Assume $p_1(Y(a) \mid w) \neq p_2(Y(a) \mid w)$, and a fixed $p_1(W),p_2(W)$ (either equal or not).  Then there exists $\tilde{p}(A \mid W)$ such that
$\sum_{w,a} p_1(Y(a) \mid w) \tilde{p}(a\mid w)p_1(w) \neq \tilde{p}_1(Y) \neq \tilde{p}_2(Y) = \sum_{w,a} p_2(Y(a) \mid w) \tilde{p}(a\mid w) p_2(w)$.
\label{lem:simple-cond}
\end{lem}

\begin{thm}
Assume $p(\{ Y({\bf a}) \mid Y \in {\bf Y}^* \})$ is not identifiable, where ${\bf Y}^* = \an_{{\cal G}_{{\bf f}_{\bf A}}}({\bf Y}) \setminus {\bf A}$.  Then
$p(\{ Y({\bf f}_{\bf A}) \mid Y \in {\bf Y} \})$ is not identifiable.
\label{thm:id-comp-f}
\end{thm}


\begin{lem}
Assume a fixed $p_1(W)$,$p_2(W)$ (equal or not), and
$p_1(Y_1(a_1,a_2), Y_2(a_1,a_2) | w) \neq p_2(Y_1(a_1,a_2), Y_2(a_1,a_2) | w)$. 
Then there exists $\tilde{p}(A_1 | W), \tilde{p}_2(A_2 | W)$ such that
{\small
\begin{align*}
\sum_{w,a_1,a_2} p_1(Y_1(a_1,a_2), Y_2(a_1,a_2) | w) \tilde{p}(a_1| w)\tilde{p}(a_2| w)p(w)
\neq\\
\sum_{w,a_1,a_2} p_2(Y_1(a_1,a_2), Y_2(a_1,a_2) | w) \tilde{p}(a_1| w)\tilde{p}(a_2| w) p(w).
\end{align*}
}
\label{lem:simple-cond-e}
\end{lem}

\begin{thm}
Let ${\bf Y}^* = \an_{{\cal G}_{{\bf f}_{{\bf A}_{\alpha}}}}({\bf Y}) \setminus {\bf A}_{\alpha}$.
Then if $p({\bf Y}^*({\mathfrak a}_{\alpha}))$ is not identifiable,
$p({\bf Y}((\tilde{\bf f}_{\bf A})_{\alpha,\overline{\alpha}}))$
is not identifiable.
\label{thm:id-comp-fe}
\end{thm}

\section*{Finding Optimal Total Effect Policies}
\label{sec:opt-total-policy}
Generally, having defined counterfactual responses to policies, the goal is to find an \emph{optimal} policy.  Optimality may be
quantified in a number of ways, but a common approach is maximizing expected counterfactual outcome under a policy, that is
finding $\argmax_{{\bf f}_{\bf A}} \E[Y({\bf f}_{\bf A})]$.

Consider a simple example in Fig. \ref{fig:triangle} (a), where we wish to learn the optimal policy $f_A(W)$, by which we mean a policy that
maximizes $\E[Y(f_A(W))]$.  Assume we happen to
know the correct parametric specification for $Q(A,W;\gamma) = \E[Y(A) \mid W; \gamma] = \E[Y | A,W; \gamma]$
(this type of specification is sometimes known as a Q-function in the reinforcement learning literature).
Then for a binary treatment $A$ (with values $0,1$), it is fairly easy to show that for any given $w \in {\mathfrak X}_W$,
the optimal policy has the form $\I(Q(1,w;\gamma) > Q(0,w;\gamma))$, where $\I(.)$ is the indicator function.


Consider a general version of Fig. \ref{fig:triangle} (b), with a vector of baseline factors $W_0$, a set of treatments $A_i$, and outcomes $W_i$ for $i = 1, \ldots, k$, and a temporal order on variables $W_0, A_1, W_1, \ldots, A_k, W_k$, where all variables earlier in the order are assumed to causally influence variables later in the order, and an unobserved common parent $U$ of $W_0, \ldots, W_k$.  Given $A_i$, denote $\overline{A}_i$ to be all variables earlier in the ordering than $A_i$, similarly for $W_i$ and $\overline{W}_i$.
Finally, denote by $\overline{A}_i^{-j}$ to be all variables earlier in the ordering than $A_i$ except $A_j$, similarly for $\overline{W}_i^{-j}$.
Assume we are interested in choosing a set of policies
${\bf f}_{\bf A} \equiv \{ f_{A_i}(\overline{A}_i) \mid i = 1, \ldots, k\}$ which maximize $\E[W_k({\bf f}_{\bf A})]$.

It's well-known 
that this counterfactual mean is identified, under the model we specified,
via a version of the g-computation algorithm \cite{robins86new}, a special case of (\ref{eqn:id-f}):
{\small
\begin{align}
\notag
\sum_{\overline{W}_k} \E[W_k \mid W_0, \ldots, W_{k-1}, f_{A_1}(W_0), \ldots, f_{A_k}(\overline{A}_k)] \times \\
\prod_{i=1}^{k-1} p(W_{i} \mid W_0, \ldots, W_{i-1}, f_{A_1}(W_0), \ldots, f_{A_i}(\overline{A}_i)).
\label{eqn:id-f-g}
\end{align}
}

If we knew the correct specification of all these models, we could obtain the optimal $f^*_{A_k}$ as
$\mathbb{I}(\E[ W_k | A_k=1, \overline{A}_k] > \E[ W_k | A_k=0, \overline{A}_k])$, and the optimal
$f^*_{A_i}$ as
{\small
\begin{align}
\notag
\mathbb{I}
[
\E[ W_k(
f^*_{A_k}, \ldots, f^*_{A_{i+1}},
A_i=1) | \overline{A}_{i} ] >\\
\E[ W_k(
f^*_{A_k}, \ldots, f^*_{A_{i+1}},
A_i=0) | \overline{A}_{i} ]
]
\label{eqn:ind-seq}
\end{align}
}
by an appropriate modification of (\ref{eqn:id-f-g}).
This approach is known as dynamic programming or backwards induction.

\subsection*{Outcome Weighted Learning}
\label{sec:owl}
If models in (\ref{eqn:id-f-g}) are not known precisely, but their parametric form is known, they may be estimated from data via usual maximum likelihood
methods, and resulting estimates used to find the optimal policy given data.  If we are uncertain whether these models are correctly specified,
we are naturally no longer guaranteed to learn optimal policies directly in this way.
In addition even under a correctly specified model, evaluating (\ref{eqn:ind-seq}) may be computationally demanding, as it involves integrating over
$W_{i}, \ldots, W_{k-1}$.

A number of alternative strategies for finding policies where the optimal policy is not easily available in closed form have been considered in the literature,
including grid search, and value search within a restricted class of policies.  In this paper, we consider outcome weighted learning \cite{chen16owl},
which translates the problem of picking optimal policies into a weighted classification problem in machine learning.

The key idea is if we had \emph{any} method for learning $\E[ W_k(f^*_{A_k}, \ldots, f^*_{A_{i+1}}, A_i) \mid \overline{A}_{i} ]$ for any $A_i$,
the task for finding $f^*_{A_i}$, assuming $f^*_{A_k}, \ldots, f^*_{A_{i+1}}$ were already found recursively, reduces to training a classifier
mapping features $\overline{A}_i$ to a class label $A^*_i$ (either $0$ or $1$).  In typical binary classification problems, training a classifier entails minimizing
the 0-1 loss, where a correct classification is not penalized, while an incorrect classification is penalized by $1$.  In our case, we solve a sequence of recursive
classification problems with a weighted 0-1 loss.  In the base case, choosing $A^*_k$ correctly yields no penalty, while choosing $A^*_k$ incorrectly yields the penalty
{\small
\begin{align*}
\E[ W_k \mid A^*_k, \overline{A}_k] - \E[ W_k \mid 1-A^*_k, \overline{A}_k].
\end{align*}
}
Assuming $f^*_{A_k}, \ldots, f^*_{A_{i+1}}$ were already selected via classifiers minimizing appropriate loss,
choosing $A^*_i$ correctly yields no penalty, while choosing $A^*_i$ incorrectly yields a penalty
{\small
\begin{align*}
\E[ W_k(f^*_{A_k}, \ldots, f^*_{A_{i+1}}, A_i=A^*_i) \mid \overline{A}_{i} ] -\\
\E[ W_k(f^*_{A_k}, \ldots, f^*_{A_{i+1}}, A_i=1-A^*_i) \mid \overline{A}_{i} ].
\end{align*}
}
A number of approaches have been developing for minimizing these types of non-convex losses.
While any choice that specifies some inductive bias may imply the true optimal policy is no longer within the considered class,
a flexible classification strategy minimizes this risk in practice.  In our simulations, we used support vector machines (SVMs)
\cite{cortes95svm}, because of their flexibility and relative simplicity.
Note that the choice of classifier implicitly defines the restricted set of policies where we seek to find the optimum.
In the case of SVMs, this choice is all policies defined by a hyperplane through a high dimensional space defined by a kernel function.
We now consider how outcome weighted learning and dynamic programming translate to optimizing path-specific policies.

\section*{Finding Optimal Path-Specific Effect Policies}
\label{sec:opt-path-pol}
Consider the generalization of Fig. \ref{fig:triangle} (b) to the longitudinal setting with mediators, shown (for two time points) in
Fig. \ref{fig:med_long} (c).  This causal model corresponds to the setting described in detail in \cite{caleb17quantifying}, representing
an observational longitudinal study of HIV patients.  Here, $W_0$ represents the baseline variables of a patient, $A_1,A_2$ represent
treatment assignments, which were chosen based on observed treatment history according to physician's best judgement, $W_1,W_2$
are intermediate and final outcomes (such as CD4 count or viral failure), and $M_1,M_2$ are measures of patient adherence to their
treatment regimen.  We are interested in finding policies $f_{A_1}(W_0), f_{A_2}(W_0,W_1,A_1,M_1)$ that optimize the effect of
$A_1,A_2$ on $W_2$ that is either direct or via intermediate outcomes, but not via adherence, and where adherence is kept to that of
a reference treatment $a_1,a_2$.  Specifically, we are interesting in choosing
$f_{A_1},f_{A_2}$ to optimize the following counterfactual expectation:
{\small
\begin{align*}
\E\left[W_2\left(
	\begin{array}{c}
	W_0,
	f_{A_1}(W_0),
	M_1(a_1),
	W_1(f_{A_1}(W_0), M_1(a_1)),\\
	M_2(a_1, a_2, W_1(f_{A_1}(W_0), M_1(a_1)), M_1(a_1)),
	\end{array}
	\right)\right],
\end{align*}
}
which is identified as
{\scriptsize
\begin{align}
\sum_{\overline{W}_2}
\!\!\!\!
\begin{array}{c}
\E[ W_2 | W_0, W_1, M_1, M_2, f_{A_1}(W_0), f_{A_2}(W_0, f_{A_1}(W_0), M_1, W_1) ]\times\\
p(W_0) p(M_1 | a_1,W_0) p(W_1 | W_0, f_{A_1}(W_0), M_1) p(M_2 | W_0, W_1, a_1, a_2)
\end{array}
\label{eqn:id-g-m}
\end{align}
}
More generally for $k$ time points, we wish to learn policies $f_{A_1}, \ldots, f_{A_k}$ that
optimize
{\small
\[
\E\left[
W_k\left(
\begin{array}{c}
\{f_{A_1}\}_{(A_1W_1)_{\to},\ldots,(A_1W_k)_{\to}},\ldots,\{f_{A_k}\}_{(A_kW_k)_{\to}},\\
(a_1)_{(A_1M_1)_{\to},\ldots,(A_1M_k)}, \ldots, (a_k)_{(A_kM_k)_{\to}}
\end{array}
\right)\right],
\]
}
which is identified by the appropriate generalization of (\ref{eqn:id-g-m}):
{\small
\begin{align}
\notag
\sum_{\overline{W}_k}
\E[ W_k | W_0,\ldots,W_{k-1},M_1,\ldots,M_k,f_{A_1}(W_0), \ldots , f_{A_k}(\overline{A}_k)] \times\\
\label{eqn:id-g-mf}
\left( \prod_{i=1}^{k-1} p(W_i | W_0,\ldots,W_{i-1},f_{A_1}(W_0),\ldots,f_{A_i}(\overline{A}_i)) \right) \times \\
\notag
\left( \prod_{i=1}^{k} p(M_i | W_0, \ldots, W_{i-1}, a_1, \ldots, a_i) \right).
\end{align}
}
The dynamic programming approach to learning optimal policies within a restricted class given
by our chosen classification method proceeds as follows.  We solve a sequence of recursive
classification problems with a weighted 0-1 loss.  In the base case, choosing $A^*_k$ correctly yields no penalty,
while choosing $A^*_k$ incorrectly yields the penalty
{\small
\begin{align*}
\sum_{M_k} \{ \E[ W_k | A^*_k, \overline{A}_k] - \E[ W_k | 1-A^*_k, \overline{A}_k] \} p(M_k | a_k, \overline{A}_k).
\end{align*}
}
Assuming $f^*_{A_k}, \ldots, f^*_{A_{i+1}}$ were already selected via classifiers minimizing appropriate loss,
choosing $A^*_i$ correctly yields no penalty, while choosing $A^*_i$ incorrectly yields a penalty
$\E[ \mathcal{R}(A^*_i) - \mathcal{R}(1 - A^*_i) \mid \overline{A}_i ]$, where $\mathcal{R}(\tilde{A}_i)$ is 
{\scriptsize
\begin{align*}
W_k\left[
\begin{array}{c}
\{f^*_{A_{i+1}}\}_{\{(A_{i+1}W_{i+1})_{\to},\ldots,(A_{i+1}W_k)_{\to}\}},\ldots,\{f^*_{A_k}\}_{(A_kW_k)_{\to}},\\
\tilde{A}_i,(a_{i+1})_{\{(A_{i+1}M_{i+1})_{\to},\ldots,(A_{i+1}M_k)\}}, \ldots, (a_k)_{(A_kM_k)_{\to}}
\end{array}
\right],
\end{align*}
}
and identified by the appropriate modification of (\ref{eqn:id-g-mf}).
In the subsequent section, we report the results of a simulation study illustrating this dynamic programming approach
with an SVM classifier, for the two time point case ($k=2$).

\section{A Simulation Study}
\label{sec:simstudy}
We demonstrated our method via a simulation study 
for the model given in
Fig. \ref{fig:med_long} (c). 
We used a 5-variate normal for $W_0$, logistic regression models for binary $A_1,M_1,A_2,M_2$, and linear regressions for
continuous-valued $W_2$ and $W_1$ (the latter was a 6-variable vector).  We will make our simulation code available upon request.
We used softmargin SVMs which allowed row weights, as implemented in the \texttt{libsvm} library, to optimize the weighted 0-1 loss,
as defined in the previous section, using an appropriate hinge loss convex surrogate.

Table \ref{tab:svm} summarizes the performance of the SVM classifiers in Stage 2 and Stage 1, respectively. We report the accuracy of the classifiers trained using both linear and polynomial (poly) kernels of various degrees (d). Table \ref{tab:weight} summarizes the average weighted 0-1 loss incurred during policy evaluation in both stages.
As expected, it is straightforward to optimize the second stage policy, since significant information is available at that point via the vector $\overline{A}_2$.
At stage 1, only $W_0$ is available, which complicates deriving an optimal decision surface.  In addition, while flexible decision surfaces are helpful at the
second stage, they appear to be counterproductive for the first stage.  We leave a detailed investigation of weighted classification algorithms for this problem
to future work.

\begin{table}[ht]
\scriptsize
\centering
\caption{\small\label{tab:svm}Summary of weighted SVM performance accuracy for 2 time points }
\vspace{0.3cm}  
\renewcommand{\arraystretch}{1.1}
\begin{tabular}{|c|c|c|c|c|c|}
\hline
\multicolumn{2}{|c|}{Stage 2} & \multicolumn{4}{|c|} {Stage 1 training accuracy ($\%$)} \bigstrut\\
\cline{3-6}
\multicolumn{2}{|c|}{ training accuracy ($\%$)} &Linear & Poly (d=3) & Poly (d=5) & Poly (d=7) \bigstrut  \\
\hline\hline
Linear & 78.69  & 83.47& 75.56 & 67.08& 67.67 \bigstrut \\
\hline
Poly (d=3)& 93.53 &82.21 & 74.15& 64.61 & 65.11 \bigstrut \\
\hline
Poly (d=5)&99.18 &82.10 & 73.75& 64.76 &64.87 \bigstrut \\
\hline
Poly (d=7)&99.30 &82.59 &74.48 & 65.15 & 66.19 \bigstrut \\
\hline
\end{tabular}
\end{table}

\begin{table}[ht]
\scriptsize
\centering
\caption{\small\label{tab:weight}Summary of average weighted 0-1 loss incurred during 2-stage policy learning.}
\vspace{0.3cm}  
\renewcommand{\arraystretch}{1.1}
\begin{tabular}{|c|c|c|c|c|c|}
\hline
\multicolumn{2}{|c|}{Stage 2 average} & \multicolumn{4}{|c|} {Stage 1 average weighted 0-1 loss (3 sig.figures)} \bigstrut\\
\cline{3-6}
\multicolumn{2}{|c|}{ weighted 0-1 loss} &Linear & Poly (d=3) & Poly (d=5) & Poly (d=7) \bigstrut  \\
\hline\hline
Linear& $3.762$ & $145$ & $252$& $508$ & $491$\bigstrut \\
\hline
Poly (d=3)& $0.302 $ & $148$ &$257$ & $520$ &  $508$ \bigstrut \\
\hline
Poly (d=5)& $0.007$ & $152$ & $261$ & $532$ & $513$ \bigstrut \\
\hline
Poly (d=7)& $0.022$ & $143$ & $248$& $532$ & $ 498$\bigstrut \\
\hline
 \end{tabular}
\end{table}

\section*{Conclusion}

In this paper, we defined counterfactual responses to policies that set treatment value in such a way that they affect outcomes
with respect to certain causal pathways only.  Such counterfactuals arise when we wish to personalize only some portion of the causal
effect of the treatment, while keeping other portions to some reference values.  An example might be optimizing the chemical effect of
a drug, while keeping drug adherence to a reference value.

We gave a general algorithm for identifying these responses from data, which generalizes similar algorithms due to \cite{tian08dynamic,shpitser13cogsci}
for dynamic treatment regimes, and path-specific effects, respectively, shown that given an unrestricted class of policies the algorithm is, in some sense,
complete, and demonstrated how path-specific policies may be optimized using outcome weighted learning \cite{chen16owl}.


\clearpage
\small
\bibliographystyle{aaai}
\bibliography{references}

\section*{Appendix}

\subsection*{Example Derivation For An Identifiable Path-Specific Policy}

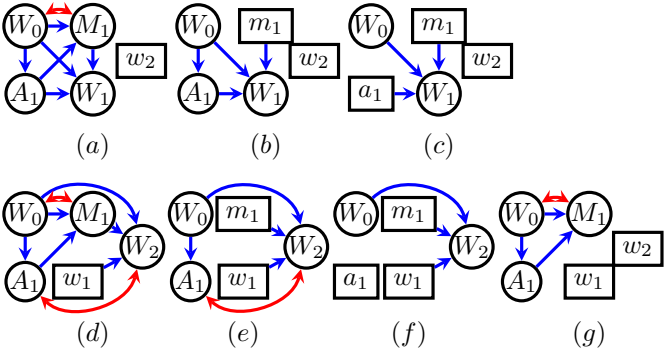
\begin{figure}
\begin{center}
\begin{tikzpicture}[>=stealth, node distance=0.9cm]
    \tikzstyle{format} = [draw, very thick, circle, minimum size=5.0mm,
	inner sep=0pt]
	\tikzstyle{square} = [draw, very thick, rectangle, minimum size=4mm]

	\begin{scope}[xshift=0.0cm]
		\path[->, very thick]
			node[format] (a1) {$A_1$}
			node[format, above of=a1] (w0) {$W_0$}
			node[format, right of=w0] (m1) {$M_1$}
			node[format, right of=a1] (w1) {$W_1$}
			node[square, above right of=w1, yshift=-0.2cm, xshift=0.2] (w2) {$w_2$}
			
			(w0) edge[blue] (a1)
			(a1) edge[blue] (m1)
			(m1) edge[blue] (w1)
			
			(w0) edge[blue] (m1)
			(w0) edge[blue] (w1)
			
			(a1) edge[blue] (w1)
			
			(w0) edge[<->, red, bend left=30] (m1)
			
			node[below of=m1, yshift=-0.7cm, xshift=0.0cm] (l) {$(a)$}
			;
	\end{scope}
	\begin{scope}[xshift=2.3cm]
		\path[->, very thick]
			node[format] (a1) {$A_1$}
			node[format, above of=a1] (w0) {$W_0$}
			node[square, right of=w0] (m1) {$m_1$}
			node[format, right of=a1] (w1) {$W_1$}
			node[square, above right of=w1, yshift=-0.2cm, xshift=0.2] (w2) {$w_2$}
			
			(w0) edge[blue] (a1)
			(m1) edge[blue] (w1)
			
			(w0) edge[blue] (w1)
			
			(a1) edge[blue] (w1)
			
			
			node[below of=m1, yshift=-0.7cm, xshift=0.0cm] (l) {$(b)$}
			;
	\end{scope}
	\begin{scope}[xshift=4.6cm]
		\path[->, very thick]
			node[square] (a1) {$a_1$}
			node[format, above of=a1] (w0) {$W_0$}
			node[square, right of=w0] (m1) {$m_1$}
			node[format, right of=a1] (w1) {$W_1$}
			node[square, above right of=w1, yshift=-0.2cm, xshift=0.2] (w2) {$w_2$}
			
			(m1) edge[blue] (w1)
			
			(w0) edge[blue] (w1)
			
			(a1) edge[blue] (w1)
			
			
			node[below of=m1, yshift=-0.7cm, xshift=0.0cm] (l) {$(c)$}
			;
	\end{scope}
	\begin{scope}[yshift=-2.5cm, xshift=0.0cm]
		\path[->, very thick]
			node[format] (a1) {$A_1$}
			node[format, above of=a1] (w0) {$W_0$}
			node[format, right of=w0] (m1) {$M_1$}
			node[square, right of=a1, xshift=-0.2cm] (w1) {$w_1$}
			node[format, below right of=m1, yshift=0.2cm, xshift=0.6] (w2) {$W_2$}
			
			(w0) edge[blue] (a1)
			(a1) edge[blue] (m1)
			(w1) edge[blue] (w2)
			
			(w0) edge[blue] (m1)
			(w0) edge[blue, bend left=60] (w2)
			
			(m1) edge[blue] (w2)
			(a1) edge[<->, red, bend right=60] (w2)
			
			(w0) edge[<->, red, bend left=30] (m1)
			
			node[below of=m1, yshift=-0.7cm, xshift=0.0cm] (l) {$(d)$}
			;
	\end{scope}
	\begin{scope}[yshift=-2.5cm, xshift=2.2cm]
		\path[->, very thick]
			node[format] (a1) {$A_1$}
			node[format, above of=a1] (w0) {$W_0$}
			node[square, right of=w0, xshift=-0.2cm] (m1) {$m_1$}
			node[square, right of=a1, xshift=-0.2cm] (w1) {$w_1$}
			node[format, below right of=m1, yshift=0.2cm, xshift=0.2cm] (w2) {$W_2$}
			
			(w0) edge[blue] (a1)
			(w1) edge[blue] (w2)
			
			(w0) edge[blue, bend left=60] (w2)
			
			(m1) edge[blue] (w2)
			(a1) edge[<->, red, bend right=60] (w2)
			
			
			node[below of=m1, yshift=-0.7cm, xshift=0.0cm] (l) {$(e)$}
			;
	\end{scope}
	\begin{scope}[yshift=-2.5cm, xshift=4.4cm]
		\path[->, very thick]
			node[square] (a1) {$a_1$}
			node[format, above of=a1] (w0) {$W_0$}
			node[square, right of=w0, xshift=-0.2cm] (m1) {$m_1$}
			node[square, right of=a1, xshift=-0.2cm] (w1) {$w_1$}
			node[format, below right of=m1, yshift=0.2cm, xshift=0.2cm] (w2) {$W_2$}
			
			(w1) edge[blue] (w2)
			
			(w0) edge[blue, bend left=60] (w2)
			
			(m1) edge[blue] (w2)
			
			
			node[below of=m1, yshift=-0.7cm, xshift=0.0cm] (l) {$(f)$}
			;
	\end{scope}
	\begin{scope}[yshift=-2.5cm, xshift=6.6cm]
		\path[->, very thick]
			node[format] (a1) {$A_1$}
			node[format, above of=a1] (w0) {$W_0$}
			node[format, right of=w0] (m1) {$M_1$}
			node[square, right of=a1] (w1) {$w_1$}
			node[square, above right of=w1, yshift=-0.2cm, xshift=0.2] (w2) {$w_2$}
			
			(w0) edge[blue] (a1)
			(a1) edge[blue] (m1)
			
			(w0) edge[blue] (m1)
			
			
			(w0) edge[<->, red, bend left=30] (m1)
			
			node[below of=m1, yshift=-0.7cm, xshift=0.0cm] (l) {$(g)$}
			;
	\end{scope}

\end{tikzpicture}
\end{center}
\caption{
For ${\cal G}$ in Fig. \ref{fig:med_long} (a):
(a) $\phi_{\{W_2\}}({\cal G})$,
(b) $\phi_{\{W_2,M_1\}}({\cal G})$,
(c) $\phi_{\{W_2,M_1,A_1\}}({\cal G})$,
(d) $\phi_{\{W_1\}}({\cal G})$,
(e) $\phi_{\{W_1,M_1\}}({\cal G})$,
(f) $\phi_{\{W_1,M_1,A_1\}}({\cal G})$,
(g) $\phi_{\{W_2,W_1\}}({\cal G})$.
}
\label{fig:med_long_fix}
\end{figure}

We seek to identify the distribution $p(W_2(\{ f_{A_1}(W_0) \}_{(A_1W_1)_{\to}}, a_{(A_1M_1)_{\to}}))$ in Fig. \ref{fig:med_long} (a).
Then ${\bf Y}^* = \{ W_2, W_1, M_1, W_0 \}$, and ${\cal D}({\cal G}_{{\bf Y}^*}) = \{ \{ W_2 \}, \{ W_0, M_1 \}, \{ W_1 \} \}$
(the graph ${\cal G}_{{\bf Y}^*}$ is shown in Fig. \ref{fig:med_long} (b)).
Thus, we have three terms, a term $\phi_{\{ W_0, M_1, A_1, W_1 \}}(p;{\cal G})$ for $W_2$, a term
$\phi_{\{W_0, A_1, M_1, W_2\}}(p;{\cal G})$ for $W_1$, and a term $\phi_{\{ A_1, W_1, W_2\}}(p;{\cal G})$ for $\{ W_0, M_1 \}$.
We have
{\small
\begin{align*}
\phi_{\{W_0, A_1, M_1, W_2\}}(p;{\cal G})
&=
\phi_{\{W_0,A_1,M_1\}}\left(\sum_{W_2} p; {\cal G}^{(a)}\right)\\
&= \phi_{\{W_0,A_1\}}\left( \frac{p(W_0,A_1,M_1,W_1)}{p(M_1 \mid A_1, W_0)}; {\cal G}^{(b)} \right)\\
&= \phi_{\{W_0\}}\left( \frac{p(W_0,A_1,M_1,W_1)}{p(M_1,A_1 \mid W_0)}; {\cal G}^{(c)} \right)\\
&= p(W_1 \mid M_1, A_1, W_0),
\end{align*}
}
where ${\cal G}^{(a)},{\cal G}^{(b)},{\cal G}^{(c)}$ are CADMGs in Figs. \ref{fig:med_long_fix} (a), (b), and (c), respectively.
Similarly, $\phi_{\{ W_0, M_1, A_1, W_1 \}}(p;{\cal G})$ is equal to
{\small
\begin{align*}
& \phi_{\{ W_0, M_1, A_1\}}\left(\frac{p(W_0,A_1,M_1,W_1,W_2)}{p(W_1 \mid M_1,A_1,W_0)}; {\cal G}^{(d)} \right)\\
&=
\phi_{\{ W_0, A_1\}}\left(\frac{p(W_0,A_1,M_1,W_1,W_2)}{p(W_1,M_1 \mid A_1,W_0)}; {\cal G}^{(e)} \right)\\
&=
\phi_{\{ W_0\}}\left(\sum_{A_1} \frac{p(W_0,A_1,M_1,W_1,W_2)}{p(W_1,M_1 \mid A_1,W_0)}; {\cal G}^{(f)} \right)\\
&= \sum_{W_0,A_1} p(W_2 \mid W_1,M_1,A_1,W_0) p(A_1,W_0),
\end{align*}
}
where ${\cal G}^{(d)},{\cal G}^{(e)},{\cal G}^{(f)}$ are CADMGs in Figs. \ref{fig:med_long_fix} (d), (e), and (f), respectively.
Finally,
{\small
\begin{align*}
\phi_{\{ A_1,W_1,W_2 \}}(p;{\cal G})
&=
\phi_{\{ A_1,W_1\}}\left( \sum_{W_2} p; {\cal G}^{(a)} \right)\\
&=
\phi_{\{A_1\}}\left( \sum_{W_2,W_1} p; {\cal G}^{(g)} \right)\\
&= \frac{p(W_0,A_1,M_1)}{p(A_1 \mid W_0)} = p(M_1 \mid A_1, W_0) p(W_0),
\end{align*}
}
where ${\cal G}^{(a)},{\cal G}^{(g)}$ are CADMGs in Figs. \ref{fig:med_long_fix} (a), and (g), respectively.
Note that whenever the fixing operation for a kernel $q_{\bf V}({\bf V} \mid {\bf W})$ that fixes $V \in {\bf V}$ is such that
${\bf V} \setminus \{ V \} \subseteq \nd_{{\cal G}({\bf V},{\bf W})}(V)$, the resulting kernel can be viewed as
$\tilde{q}_{{\bf V} \setminus \{V\}}({\bf V}\setminus\{V\}\mid{\bf W}\cup\{V\}) = \sum_V q_{\bf V}({\bf V} \mid {\bf W})$.
We now combine these terms, evaluating $A_1$ to either $a$ or $f_{A_1}(W_0)$, as appropriate, yielding the
following expression for
$p(W_2(\{ f_{A_1}(W_0) \}_{(A_1W_1)_{\to}}, a_{(A_1M_1)_{\to}}))$:
{\small
\begin{align*}
\sum_{W_0,A_1,M_1,W_1}
\!\!\!\!\!
\!\!\!\!\!
\begin{array}{c}
\left[
p(W_1 \mid M_1, f_{A_1}(W_0), W_0)
\right]
\left[
p(M_1 \mid a_1, W_0) p(W_0)
\right]\times\\
\left[
\sum_{W_0,A_1} 
p(W_2 \mid W_1,M_1,A_1,W_0) p(A_1,W_0)
\right].
\end{array}
\end{align*}
}

\subsection*{Proofs}

\begin{thma}{\ref{thm:id-e}}
Under above assumptions,
$p(\{ {\bf V} \setminus {\bf A}_{\alpha} \}({\mathfrak a}_{\alpha}))$ is 
{\small
\begin{align}
\sum_{{\bf Y}^* \setminus {\bf Y}} \prod_{{\bf D} \in {\cal D}({\cal G}_{{\bf Y}^*})}
\phi
^{{\mathfrak a}_{ \{ (AD)_{\to} \in \alpha \mid D \in {\bf D}, A \not\in {\bf D} \} }}
_{{\bf V} \setminus {\bf D}}(p({\bf V}); {\cal G}).
\end{align}
}
\end{thma}
\begin{prf}
Assume an NPSEM-IE for a DAG ${\cal G}({\bf H}\cup{\bf V})$.  Then for any disjoint ${\bf S}_1,{\bf S}_2 \subseteq {\bf V}$,
$\{ S_1({\bf a}_1) \mid S_1 \in {\bf S}_1, {\bf a}_1 \in {\mathfrak X}_{\pa^s_{{\cal G}({\bf V})}({\bf S}_1)} \}$ is marginally independent of
$\{ S_2({\bf a}_2) \mid S_2 \in {\bf S}_2, {\bf a}_2 \in {\mathfrak X}_{\pa^s_{{\cal G}({\bf V})}({\bf S}_2)} \}$ if ${\bf S}_1$ is marginally
d-separated in the graph ${\cal G}_{({\bf H} \cup {\bf V}) \setminus (\pa^s_{{\cal G}({\bf V})}({\bf S}_1) \cup \pa^s_{{\cal G}({\bf V})}({\bf S}_1))}$.
Note that this is true even if $\pa^s_{{\cal G}({\bf V})}({\bf S}_1) \cap \pa^s_{{\cal G}({\bf V})}({\bf S}_2) \neq \emptyset$, and
${\bf a}_1$ and ${\bf a}_2$ assign different values to variables in this intersection.
Pick a value assignment ${\bf v} \in {\mathfrak X}_{\bf V}$. Under our assumptions,
$p(\{ {\bf V} \setminus {\bf A}_{\alpha} \}({\mathfrak a}_{\alpha}))$
is equal to
{\small
\[
\prod_{{\bf D} \in {\cal D}({\cal G}_{{\bf Y}^*})}
p({\bf D}({\bf a}_{\pa^s_{{\cal G}({\bf V})}({\bf D}) \cap {\bf A}},{\bf v}_{\pa^s_{{\cal G}({\bf V})}({\bf D}) \setminus {\bf A}}) = {\bf v}_D),
\]
}
where ${\bf a}_{\pa^s_{{\cal G}({\bf V})}({\bf D}) \cap {\bf A}}$ is the set of values of ${\bf A} \cap \pa^s_{{\cal G}({\bf V})}({\bf D}$
assigned by ${\mathfrak a}$ to edges between that set and ${\bf D}$.
A detailed derivation, via an argument that ``unrolls'' the counterfactuals can be found in the supplement in \cite{shpitser13cogsci}.
Each factor in the above expression is now identified via the corresponding factor in (\ref{eqn:id-e}) by Theorem 60 in \cite{richardson17nested}.
This proves the claim.
\end{prf}

\begin{thma}{\ref{thm:id-ef}}
Define ${\cal G}_{{\bf f}_{{\bf A}_{\alpha}}}$ as ${\cal G}_{{\bf f}_{\bf A}}$ before, and let
${\bf Y}^* \equiv \an_{{\cal G}_{{\bf f}_{{\bf A}_{\alpha}}}}({\bf Y}) \setminus {\bf A}_{\alpha}$.
Then
$p({\bf Y}((\tilde{\bf f}_{\bf A})_{\alpha,\overline{\alpha}}))$
is identified if
$p({\bf Y}^*({\bf a}))$
is identified, and for every ${\bf D} \in {\cal D}({\cal G}_{{\bf Y}^*})$,
and every $A \in {\bf A}_{\alpha}$, either every directed edge out of $A$ into ${\bf D}$ is in $\alpha$,
or every directed edge out of $A$ into ${\bf D}$ is not in $\alpha$.  Moreover, the identifying formula is
{\small
\begin{align}
\sum_{({\bf Y}^* \cup {\bf A}_{\alpha}) \setminus {\bf Y}} \prod_{{\bf D} \in {\cal D}({\cal G}_{{\bf Y}^*})}
	\phi^{\tilde{\bf a}_{\pa^s_{\cal G}({\bf D}) \cap {\bf A}_{\alpha}}}_{{\bf V} \setminus {\bf D}}(p({\bf V});{\cal G}),
\label{eqn:id-ef}
\end{align}
}
where ${\tilde{\bf a}_{\pa^s_{\cal G}({\bf D}) \cap {\bf A}_{\alpha}}}$ is defined to be
$\{ A = f_A({\bf W}_A) \mid A \in \pa^{\alpha}_{\cal G}({\bf D}) \cap {\bf A}_{\alpha} \} \cup
\{ A = f'_A({\bf W}_A) \mid A \in \pa^{\overline{\alpha}}_{\cal G}({\bf D}) \cap {\bf A}_{\alpha} \} $,
if $\pa_{\cal G}({\bf D}) \cap {\bf A}_{\alpha}$
is not empty, and is defined to be the empty set otherwise.
\end{thma}\\
\begin{prf}
Let $\alpha^*$ be the union of $\alpha$ and $\bar{\alpha}$.
Fix any edge intervention ${\mathfrak a}_{\alpha^*}$ such that all edges in $\bar{\alpha}$ are assigned to appropriate values of ${\bf a}$.
Then by Theorem \ref{thm:id-e},
$p({\bf Y}^*({\mathfrak a}_{\alpha^*}))$ is 
{\small
\begin{align}
\prod_{{\bf D} \in {\cal D}({\cal G}_{{\bf Y}^*})}
\phi
^{{\mathfrak a}_{ \{ (AD)_{\to} \in \alpha \mid D \in {\bf D}, A \not\in {\bf D} \} }}
_{{\bf V} \setminus {\bf D}}(p({\bf V});{\cal G}),
\label{eqn:inter}
\end{align}
}
for any ${\mathfrak a}_{\alpha^*}$.
Since, by definition,
$p({\bf Y}^*((\tilde{\bf f}_{\bf A})_{\alpha,\overline{\alpha}}))$
is equal to
$p({\bf Y}^*(\tilde{\mathfrak a}_{\alpha^*}))$,
where $\tilde{\mathfrak a}_{\alpha^*}$ sets
values of every $A$ with respect to edges in $\alpha$ via $f_A$ evaluated at
${\bf W}_A(\tilde{\mathfrak a}_{\alpha^*})$, and sets values of every $A$ with respect to edges in
$\bar{\alpha}$ similarly via $f'_A$, our conclusion follows immediately.
\end{prf}

\begin{lema}{\ref{lem:simple-cond}}
Assume $p_1(Y(a) \mid w) \neq p_2(Y(a) \mid w)$, and a fixed $p_1(W),p_2(W)$ (either equal or not).  Then there exists $\tilde{p}(A \mid W)$ such that
$\sum_{w,a} p_1(Y(a) \mid w) \tilde{p}(a\mid w)p_1(w) \neq \tilde{p}_1(Y) \neq \tilde{p}_2(Y) = \sum_{w,a} p_2(Y(a) \mid w) \tilde{p}(a\mid w) p_2(w)$.
\end{lema}\\
\begin{prf}
Since $\tilde{p}_1(Y)$ and $\tilde{p}_2(Y)$ are weighted averages, to assure their inequality it suffices to pick $\tilde{p}(A \mid W)$ in such a way that
$\tilde{p}(a \mid w) p_1(w)$ and $\tilde{p}(a \mid w) p_2(w)$ are both sufficiently close to $1$.
\end{prf}

\begin{thma}{\ref{thm:id-comp-f}}
Assume $p(\{ Y({\bf a}) \mid Y \in {\bf Y}^* \})$ is not identifiable, where ${\bf Y}^* = \an_{{\cal G}_{{\bf f}_{\bf A}}}({\bf Y}) \setminus {\bf A}$.  Then
$p(\{ Y({\bf f}_{\bf A}) \mid Y \in {\bf Y} \})$ is not identifiable.
\end{thma}\\
\begin{prf}
Order variables in ${\bf Y}^*$ topologically as $Y_1, \ldots, Y_k$ with $\overline{Y}_i$ being the set of variables in ${\bf Y}^*$ earlier in the ordering than $Y_i$.
Pick the earlier $Y_i$ in the ordering such that $p(Y_i({\bf a}) \mid \overline{Y}_i({\bf a}))$ is not identified.
Such a variable is guaranteed to exist, since $p(\{ Y({\bf a}) \mid Y \in {\bf Y}^* \})$ is not identifiable.
If $\overline{Y}_i = \emptyset$, our conclusion is trivial since $p(Y_i({\bf a})) = p(Y_i({\bf f}_{\bf A}))$ is not identifiable.  Otherwise,
by a simple extension of Lemma \ref{lem:simple-cond},
$p(\{ Y_j({\bf f}_{\bf A}) \mid Y_j \in {\overline Y}_{i+1} \setminus {\bf W}_A \})$ is not identified, and thus, neither is
$p(\{ Y_i({\bf f}_{\bf A}) \mid Y_i \in {\bf Y}^* \})$.
\end{prf}


\begin{lema}{\ref{lem:simple-cond-e}}
Assume a fixed $p_1(W)$,$p_2(W)$ (equal or not), and
$p_1(Y_1(a_1,a_2), Y_2(a_1,a_2) | w) \neq p_2(Y_1(a_1,a_2), Y_2(a_1,a_2) | w)$. 
Then there exists $\tilde{p}(A_1 | W), \tilde{p}_2(A_2 | W)$ such that
{\small
\begin{align*}
\sum_{w,a_1,a_2} p_1(Y_1(a_1,a_2), Y_2(a_1,a_2) | w) \tilde{p}(a_1| w)\tilde{p}(a_2| w)p(w)
\neq\\
\sum_{w,a_1,a_2} p_2(Y_1(a_1,a_2), Y_2(a_1,a_2) | w) \tilde{p}(a_1| w)\tilde{p}(a_2| w) p(w).
\end{align*}
}
\end{lema}\\
\begin{prf}
Since $\tilde{p}_1(Y_1,Y_2)$ and $\tilde{p}_2(Y_1,Y_2)$ are weighted averages, to assure their inequality it suffices to pick $\tilde{p}(A_1 \mid W)$,$\tilde{p}(A_2 \mid W)$
such that
$\tilde{p}(a_1 \mid w) \tilde{p}(a_2 \mid w) p_1(w)$ and $\tilde{p}(a_1 \mid w) \tilde{p}(a_2 \mid w) p_2(w)$ are both sufficiently close to $1$.
\end{prf}

\begin{thma}{\ref{thm:id-comp-fe}}
Assume
$p({\bf Y}^*({\mathfrak a}_{\alpha}))$ is not identifiable, and let ${\bf Y}^* = \an_{{\cal G}_{{\bf f}_{{\bf A}_{\alpha}}}}({\bf Y}) \setminus {\bf A}_{\alpha}$.
Then
$p({\bf Y}((\tilde{\bf f}_{\bf A})_{\alpha,\overline{\alpha}}))$
is not identifiable.
\end{thma}\\
\begin{prf}
Order variables in ${\bf Y}^*$ topologically as $Y_1, \ldots, Y_k$ with $\overline{Y}_i$ being the set of variables in ${\bf Y}^*$ earlier in the ordering than $Y_i$.
Pick the earlier $Y_i$ in the ordering such that $p(Y_i({\mathfrak a}_{\alpha}) \mid \overline{Y}_i({\mathfrak a}_{\alpha}))$ is not identified.
Such a variable is guaranteed to exist, since
$p({\bf Y}^*({\mathfrak a}_{\alpha}))$
is not identifiable.
If $\overline{Y}_i = \emptyset$, our conclusion is trivial since $p(Y_i({\mathfrak a}_{\alpha})) = p(Y_i((\tilde{\bf f}_{\bf A})_{\alpha,\overline{\alpha}}))$ is not identifiable.
Otherwise, by a simple extension of Lemma \ref{lem:simple-cond-e},
$p(\{ {\overline Y}_{i+1} \setminus {\bf W}_A \}((\tilde{\bf f}_{\bf A})_{\alpha,\overline{\alpha}}))$
is not identified, and thus, neither is
$p({\bf Y}^*((\tilde{\bf f}_{\bf A})_{\alpha,\overline{\alpha}}))$.
\end{prf}

\end{document}